\documentclass[
 reprint,
 amsmath,amssymb,
 aps,
 pra,
floatfix,
superscriptaddress, 
]{revtex4-2}

\usepackage{graphicx}
\usepackage{dcolumn}
\usepackage{bm}
\usepackage[load=physical,load=synchem]{siunitx}

\begin{document}

\newcommand{\Vector}[1]{\vec{\textbf{\textit{#1}}}}

\newcommand{\CompVec}[2]{\vec{\textbf{\textit{#1}}}_{\textbf{\textit{#2}}}}

\renewcommand{\figurename}{\textbf{Figure}}

\title{Optical Near-Field Electron Microscopy}

\author{Raphaël Marchand}
  \affiliation{University of Vienna, Faculty of Physics, VCQ, 
A-1090 Vienna, Austria}
  \affiliation{University of Vienna, Max Perutz Laboratories, 
Department of Structural and Computational Biology, A-1030 Vienna, 
Austria}
\author{Radek Šachl}
\affiliation{J. Heyrovský Institute of Physical Chemistry of the Czech Academy of Sciences, Dolejškova 3, 182 23 Prague, Czech Republic}
\author{Martin Kalbáč}
\affiliation{J. Heyrovský Institute of Physical Chemistry of the Czech Academy of Sciences, Dolejškova 3, 182 23 Prague, Czech Republic}
\author{Martin Hof}
\affiliation{J. Heyrovský Institute of Physical Chemistry of the Czech Academy of Sciences, Dolejškova 3, 182 23 Prague, Czech Republic}
\author{Rudolf Tromp}
\affiliation{IBM T.J. Watson Research Center, 1101 Kitchawan Road, Yorktown Heights, New York 10598, USA}
\affiliation{Huygens-Kamerlingh Onnes Laboratory, Leiden Institute of Physics, Leiden University, Leiden, The Netherlands}
\author{Mariana Amaro}
\affiliation{J. Heyrovský Institute of Physical Chemistry of the Czech Academy of Sciences, Dolejškova 3, 182 23 Prague, Czech Republic}
\author{Sense J. van der Molen}
\affiliation{Huygens-Kamerlingh Onnes Laboratory, Leiden Institute of Physics, Leiden University, Leiden, The Netherlands}
\author{Thomas Juffmann}

\affiliation{University of Vienna, Faculty of Physics, VCQ, 
A-1090 Vienna, Austria}
\affiliation{University of Vienna, Max Perutz Laboratories, 
Department of Structural and Computational Biology, A-1030 Vienna, 
Austria}

\begin{abstract}

Imaging dynamical processes at interfaces and on the nanoscale is of great importance throughout science and technology. While light-optical imaging techniques often cannot provide the necessary spatial resolution, electron-optical techniques damage the specimen and cause dose-induced artefacts. Here, Optical Near-field Electron Microscopy (ONEM) is proposed, an imaging technique that combines non-invasive probing with light, with a high spatial resolution read-out via electron optics. Close to the specimen, the optical near-fields are converted into a spatially varying electron flux using a planar photocathode. The electron flux is imaged using low energy electron microscopy, enabling label-free nanometric resolution without the need to scan a probe across the sample. The specimen is never exposed to damaging electrons. 

\setlength{\parindent}{0cm}

\end{abstract}

\maketitle

\section{Introduction}

Interfaces are of utmost importance throughout science, technology, industry, biology, and medicine. Imaging dynamical processes happening at interfaces can yield crucial information on the underlying processes, ranging from electroplating, to corrosion, to protein dynamics in lipid bilayers. Despite great progress over the last decades, there is currently no microscopy technique that can image dynamics at interfaces with nanometric resolution, label-free, damage-free, and over extended periods:

Optical super-resolution microscopy  ~\cite{hell1994breaking, doi:10.1021/acs.chemrev.7b00218, rust2006sub, betzig2006imaging, Dertinger22287, Gustafsson2016} has shown remarkable results over the last two decades, offering a resolution in the single digit nanometer range for selected applications (see e.g.~\cite{Sigal880, Sahl2017} and references therein). But, besides phototoxicity, fluorescence-based methods face a trade-off between frame rate, accuracy and observation time, due to the finite number of photons that can be collected per fluorophore. And while high labelling density can affect biological function~\cite{doi:10.1146/annurev-physiol-022516-034055}, low labelling density can lead to severe statistical artefacts due to blinking and bleaching~\cite{baumgart2018we}. Label-free optical techniques (such as interferometric scattering microscopy (iSCAT)~\cite{piliarik2014direct, young2018quantitative} or plasmonics enhanced protein characterization~\cite{gordon2019biosensing}) enable the detection and weighing of proteins, but their spatial resolution is diffraction limited. Resolving this issue by decreasing the wavelength to the X-ray regime is not an option for dynamic single molecule studies, as they are precluded by the trade-off between signal to noise ratio and damage~\cite{henderson1995potential}. 

Scanning probe techniques enable atomic resolution in surface imaging \cite{PhysRevLett.50.120}, and sub-nanometer resolution in imaging membrane bound proteins~\cite{engel2008structure}, but can perturb soft membranes in high speed imaging~\cite{ando2018high}. Also, the scanning of the probe can limit imaging frame rates and field of view. Similar trade-offs are faced in near-field scanning optical microscopes, where additionally the finite probe size often limits the spatial resolution to tens of nanometers~\cite{YONG2018106}.

Electron optical techniques enable the determination of the ensemble-averaged atomic structure of proteins~\cite{nogales2016development}, and the imaging of the proteome of a cell~\cite{mahamid2016visualizing}.
However, these techniques require frozen samples, which precludes real-time imaging of dynamics. Recently, thin liquid cells have enabled the observation of proteins in their native environment within electron microscopes with nanometer resolution~\cite{mirsaidov2012imaging}.
However, dose-induced damage leads to artefacts in electrochemical studies and limits extended dynamical studies of sensitive materials \cite{Woehl2020}. 
Recently, Transmission Electron Microscopy at eV-energies (eV-TEM) was developed by some of us \cite{geelen2019nonuniversal}, to decrease electron-induced damage while imaging. In fact, eV-TEM has been combined with Low Energy Electron Microscopy (LEEM) so as to image samples in electron transmission and reflection at energies of $0-\SI{100}{eV}$ \cite{geelen2019nonuniversal}. Although first results of the combination of these techniques are promising, this method is yet to prove itself for dynamical processes. 

Techniques that correlate light and electron microscopy are promising alternatives to the techniques discussed above~\cite{10.3389/fphy.2020.00047}.
Traditionally, light microscopy is first used for live imaging, or for imaging with fluorescence-enabled specificity, and electron microscopy is then used to retrieve one final high resolution snapshot of the specimen under study.

Here, we propose a novel technique that combines non-invasive probing with light with a read-out based on electron optics offering nanometric resolution. Probing and read-out are coupled via a photocathode placed in the optical near-field of the scattered light, where resolution is not limited by the optical wavelength. We will first describe the idea in more detail, then discuss its technological feasibility, its theoretical resolution and contrast, and lastly explore the potential application space of Optical Near-field Electron Microscopy (ONEM).

\section{Concept}

\begin{figure}
  \includegraphics[width=5 cm]{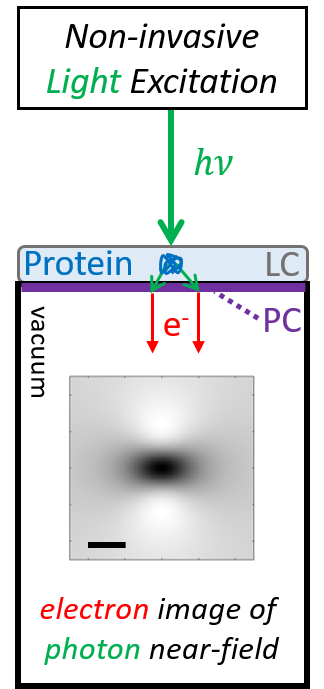}
  \caption{Optical Near-field Electron Microscopy (ONEM). Visible light illuminates a specimen (e.g. protein) in a liquid cell (LC). The resulting optical near-field interference pattern is converted into a spatially varying electron flux via a photocathode (PC). The spatial information is retrieved using aberration-corrected electron microscopy. Scalebar: $\SI{5}{nm}$. }
  \label{fig1}
\end{figure}

The proposed method is sketched in Figure \ref{fig1}: Visible light is focused onto a sample (e.g. a protein), which is close to an ultra-thin vacuum-liquid interface (e.g. in a liquid cell (LC)~\cite{de2011electron, textor2018strategies}). The sample scatters light elastically, leading to nanometric features in the near-field. This spatial light distribution is then converted, still in the near-field, into a spatially varying electron flux via the photoelectric effect within a thin layer of low-work-function photocathode (PC) material~\cite{yamaguchi2018free, yuan2015engineering}. The emitted photoelectrons are then imaged using aberration-corrected low-energy electron optics~\cite{tromp2010new, tromp2013new}. The electrons therefore provide a non-destructive read-out of the nanometric optical near-fields.

The proposed technique would be intrinsically damage free: First, low-work-function photocathodes can be efficiently excited with green light~\cite{yamaguchi2018free}, and thus at photon energies significantly below the absorption band of most proteins. The inset in Figure \ref{fig1} shows the simulated near-field intensity distribution caused by a $\SI{50}{kDa}$ protein in water and at $\SI{5}{nm}$ distance from the photocathode. Near-field simulations predict a signal to background ratio of about $\SI{1.4} {\%}$ (Figure 1, simulated using the MNPBEM toolbox~\cite{hohenester2012mnpbem}). Shot-noise limited detection, in which the intensity variance is proportional to the background, therefore requires about $5\times10^{3}$ photoelectrons per spatially resolved area to achieve a signal to noise ratio of 1. Assuming a photoelectron conversion efficiency of $\SI{3}{\%}$ (A conversion efficiency of $\approx \SI{15}{\%}$ was measured for a $25-\SI{30}{nm}$ thick photocathode \cite{yamaguchi2018free} with excitation at $\SI{2.5}{eV}$), and unity detection efficiency, this will require illuminating the protein with about $4.6\times10^{5}$ photons. For a frame rate of $\SI{1}{kHz}$ and $\SI{5}{nm}$ resolution, the required illumination intensity is about $\SI{2.5}{\micro W \, \micro m^{-2}}$, i. e. well below intensities reported in e.g. interferometric scattering microscopy (iSCAT) using light at $\SI{405} {nm}$~\cite{piliarik2014direct}. This is not surprising: Both iSCAT and ONEM are shot-noise limited. While iSCAT detects scattered light more efficiently (no photocathode), this is by far compensated by the narrower point spread function in ONEM, which increases contrast and resolution.

Second, the photocathode material will have a work function much lower than the work function of its liquid cell support ($\SI{1.9} {eV}$ for bialkali antimonide K\textsubscript{2}CsSb~\cite{yamaguchi2018free}, $\SI{1.3}{eV}$ for cesiated graphene~\cite{yuan2015engineering}, versus $\SI{4.4} {eV}$ for a few-layer graphene support~\cite{barrett2012dark}). The photocathode will be excited with visible light (e.g. $h\nu = \SI{2.3} {eV}$ for $\lambda = \SI{532} {nm}$, where $h$ is the Planck's constant, $\nu$ the frequency of the excitation light, and $\lambda$ its wavelength in vacuum), such that the energy of the created photoelectrons will be insufficient to overcome the internal potential barrier. The sample is therefore not exposed to electrons.

\section{Feasibility}

The feasibility of ONEM relies on three technologies that have recently been developed independently. 

\begin{figure}
  \includegraphics[width=\linewidth]{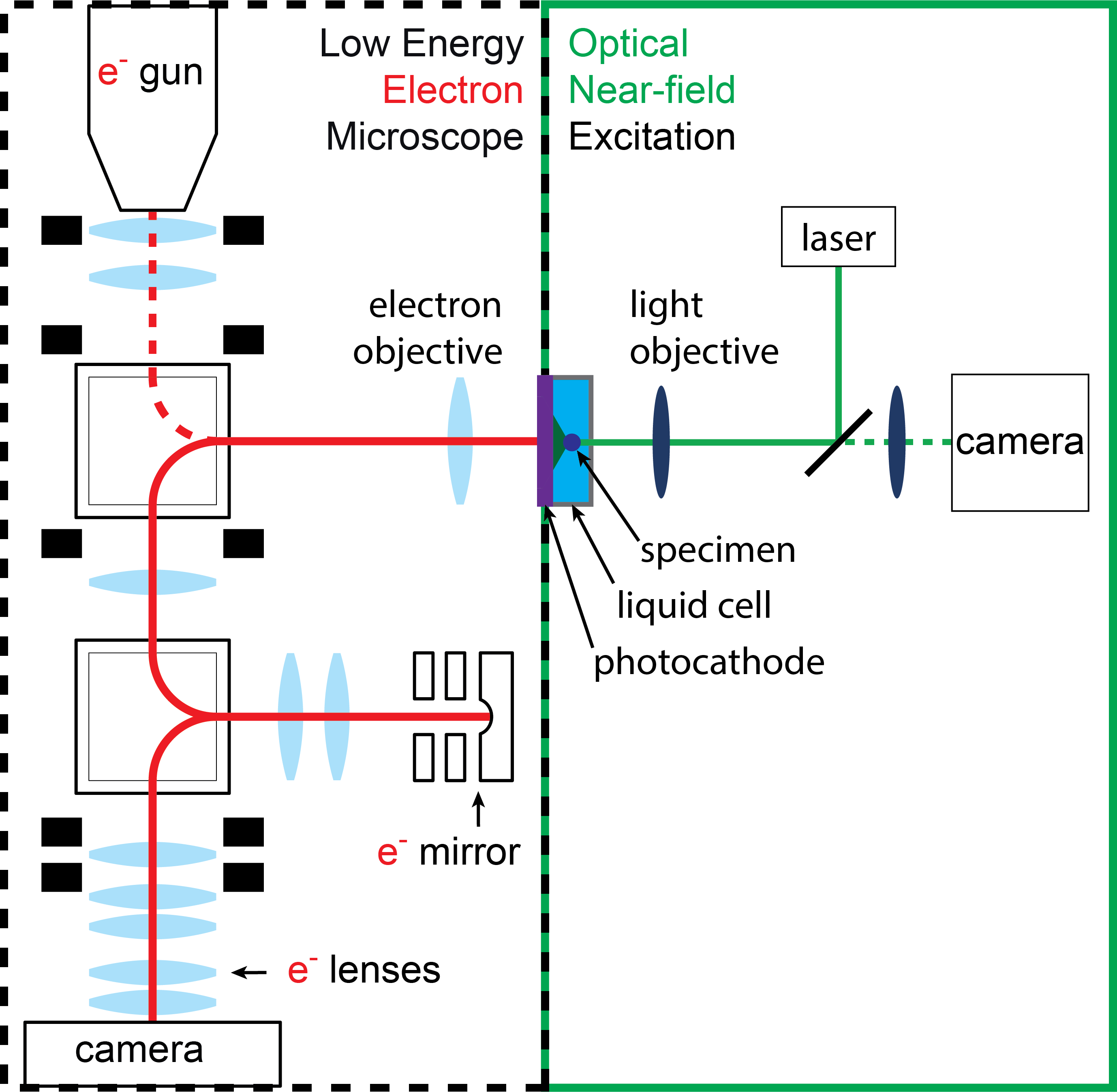}
  \caption{Schematic ONEM setup. A specimen within a liquid cell is illuminated (solid green line) and optically inspected (dashed green line) with a Continuous Wave (CW) laser. Photoelectrons (solid red line) generated in the photocathode at the backside of the liquid cell are imaged in an aberration-corrected LEEM. For this, the electrons travel through a set of lenses and are reflected by an aberration-correcting mirror. The electron gun at the top is off in ONEM, but it can be turned on to perform LEEM experiments, e. g. on the photocathode material (dashed red line).}
  \label{fig2}
\end{figure}

First, electrons emitted from the photocathode have to be imaged with nanometer resolution. Interestingly, similar requirements are to be met in LEEM~\cite{tromp2010new, tromp2013new} and eV-TEM \cite{geelen2019nonuniversal}. Typical LEEM systems have a lateral resolution of about $\SI{5} {nm}$. In the last decade however, aberration-corrected Low Energy Electron Microscopy (AC-LEEM) has been introduced, with an optimal resolution down to $\SI{1.5} {nm}$, and a field of view up to $\SI{75} {\micro m}$~\cite{tromp2010new, tromp2013new}. Moreover, using novel algorithms, many images can now be stitched together smoothly. Thus, the effective field of view can be dramatically extended, without loss of resolution~\cite{de2020quantitative}. Furthermore, we note that a resolution smaller than $\SI{3} {nm}$ has been predicted for Photo Electron Emission Microscopy (PEEM) on such state-of-the-art systems \cite{tromp2010new}. Figure \ref{fig2} shows how ONEM can be implemented within an existing LEEM setup. The LEEM sample holder will be modified to allow for optical rear-illumination of the sample. The low work-function photocathode faces the LEEM objective lens, which extracts the emitted photoelectrons in a strong electrostatic field of $\SI{100}{kV \, cm^{-1}}$. The electrons are accelerated to the column potential of $\SI{15} {keV}$, after which they follow the red solid path in Figure \ref{fig2}, from the photocathode, via an electron mirror (for aberration correction) and electron lenses, to the camera. The use of miniature objectives~\cite{Yang:16} within the light illumination path would allow for a correlative read-out using light optics (dashed green line), which could provide additional information (e.g. molecular specificity via fluorescent labelling).  

Second, in order to reach sub-optical-wavelength resolution the distance of the object to the photocathode layer needs to be much shorter than the optical wavelength. Ultra-thin, highly efficient photocathodes~\cite{yamaguchi2018free, yuan2015engineering} have just been developed, and have been shown to be smooth on the nanometer scale~\cite{feng2017near}. Most importantly, these photocathodes are efficient at green light excitation. Since these photocathodes are not to be exposed to air, they have to be prepared \textit{in situ}, e.g. via evaporation, sputter deposition, or pulsed laser deposition~\cite{doi:10.1063/1.98366} from a solid photocathode target~\cite{doi:10.1063/1.5010950}. Alternatively, alkali metals or other photocathode materials could be stabilized in between graphene layers~\cite{yamaguchi2018free,doi:10.1021/jp0377883}, potentially allowing for an \textit{ex situ} preparation of the required photocathode. 

Third, the study of specimens in their natural environment seems incompatible with the vacuum requirements for electron optics and the operation of photocathodes. Recently however, this has become possible by the use of ultrathin interfaces between vacuum and air or liquid environments~\cite{williamson2003dynamic, han2016atomically}. The latter have been developed for liquid cell TEM~\cite{de2011electron}, and modalities for electrochemistry experiments, photoactivation of specimens, and \textit{in situ} specimen mixing are now commercially available. This toolbox can be directly applied for ONEM. Note that ONEM liquid cells do not need to be ultra-thin, as it is light (and not electrons) that passes through the liquid, potentially allowing for more elaborate and robust sample manipulation.

The advance of the three techniques described above makes ONEM technologically feasible, allowing for damage-free measurements on dynamic processes in a liquid. Next, we discuss the contrast and resolution that ONEM could deliver.

\section{Contrast and Resolution }

\begin{figure}
  \includegraphics[width=\linewidth]{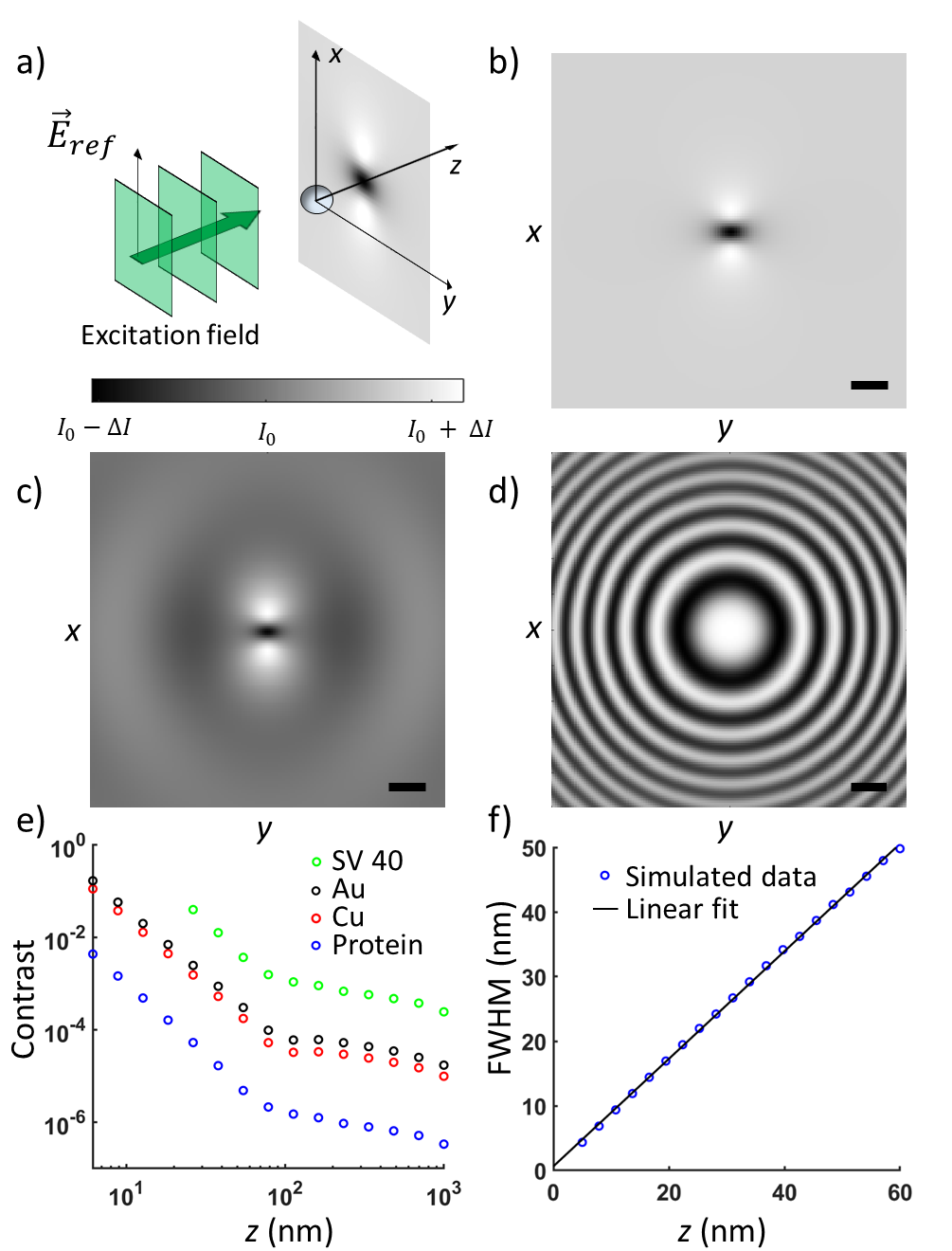}
  \caption{a) 3-dimensional drawing of a spherical nanoparticle and the interference pattern $I(x, y)$ obtained on a screen (photocathode) in the near-field ($z=\SI{5}{nm}$). b-d) simulated interference pattern $I(x, y)$ for a spherical protein of radius $R=\SI{2.5}{nm}$, and index of refraction $n_{p}=1.44$, at: b) $z=\SI{5}{nm}$ (scale bar: $\SI{10}{nm}$, $\Delta I / I_{0}=8 \times 10^{-3}$), c) $z=\SI{50}{nm}$ (scale bar: $\SI{100}{nm}$, $\Delta I / I_{0}=6 \times 10^{-6}$), d) $z=\SI{5}{\micro m}$ (scale bar: $\SI{1}{\micro m}$, $\Delta I / I_{0}=8 \times 10^{-8}$). e) simulated Michelson contrast $C=(I_{max}-I_{min})(I_{max}+I_{min})^{-1}$ of the interference pattern $I(x, y)$ as a function of the distance of the screen, for gold (Au) particles, copper (cu) particles, proteins (all $R=\SI{2.5}{nm}$), as well as for a virus (SV 40, $R=\SI{22.5}{nm}$). f)  Full Width at Half Maximum (FWHM) of the central feature of the simulated interference pattern $I(x, y)$ for a particle of radius $R=\SI{2.5}{nm}$, as a function of the distance $z$ to the particle (the FWHM is found to be independent of the particle material). 
  } 
  
  \label{fig3}
\end{figure}

In the following section we will simulate the expected contrast and resolution of ONEM. All simulations were calculated using the MNPBEM toolbox~\cite{hohenester2012mnpbem}. We assume that the sample is excited with an $x$-polarized plane wave travelling in the $z$-direction (reference field): $\CompVec{E}{ref} (\Vector{r},t)=E_{ref} \,  \textrm{e}^{i(kz - \omega t)} \CompVec{u}{x}$, where $k=2 \, \pi \, n _{m} \, \lambda ^{-1}$, with $\lambda = \SI{532}{nm} $ the wavelength of the reference field in vacuum, $\omega$ its angular frequency, $ n_{m} = 1.33 $ the refractive index of water, and $\CompVec{u}{x}$ a unitary vector in the x-direction. The particle of interest (e.g. protein, gold nanoparticle, copper cluster,…) is assumed to be spherical, with its center at the origin of the coordinate system (Figure \ref{fig3}a). 
The intensity distribution $I(\Vector{r} \,)$ results from interfering the scattered field $\CompVec{E}{scat}(\Vector{r}, t)$ with the reference field $\CompVec{E}{ref}(\Vector{r}, t)$: $I( \Vector{r} \,) \propto ||\CompVec{E}{tot}(\Vector{r}, t)|| ^{2} = ||\CompVec{E}{ref}(\Vector{r}, t)+\CompVec{E}{scat}(\Vector{r}, t)||^{2} $.
Figure \ref{fig3}b, \ref{fig3}c, and \ref{fig3}d show the intensity distribution obtained for a protein in water (radius $R=\SI{2.5} {nm}$, index of refraction $ n_{p} = 1.44$) at a distance of $z= \SI{5}{nm}$, $z= \SI{50}{nm}$, and $z= \SI{5}{\micro m}$, respectively.

On the z-axis ($y=x=0$), we observe the following behaviour: In the near-field region ($R < z \ll \lambda$), (Figure \ref{fig3}a, \ref{fig3}b, and \ref{fig3}c), the scattered field is out of phase with the reference field, resulting in a total on-axis intensity, $I(0, 0, z)$, smaller than the intensity of the reference field alone $ I_{0} \propto E_{ref}^{2}$. In the far field ($\lambda \ll z$, Figure \ref{fig3}d), the scattered field is in phase with the reference field, leading to $I(0, 0, z) > I_{0}$.

These results can be understood considering the model of an ideal radiating dipole, excited by the reference field. In this case, the total dipole moment of the particle is given by $\Vector{p} \, (t)= \epsilon_{m} \alpha \, \CompVec{E}{ref} (\Vector{0}, t) $, where $\epsilon_{m}$ is the permittivity of the surrounding medium, and $\alpha$ is the complex polarizability of the particle, given by $\alpha = 3V \frac{\epsilon_{1}-\epsilon_{m}}{\epsilon_{1}+2\epsilon_{m}}$, where $V$ is the volume of the particle, and $\epsilon_{1}$ the complex permittivity of the particle (see e. g. ch. 5.2 in \cite{bohren1998absorption}). For a protein in water, $\alpha$, $\epsilon_{1}$, and $\epsilon_{m}$ are real, and the dipole moment is in phase with the excitation field, which is consistent with the model of a Lorentz oscillator driven at frequencies far below the (material dependant) resonance frequency (ch. 3.5 in \cite{hecht2017optics}). The on-axis ($x=0$, $y=0$) scattered field is given by $\CompVec{E}{scat}(0, 0, z, t)=\frac{p (t) \, \textrm{e}^{i(kz-\omega t)}}{4\pi\epsilon_{m}}(\frac{k^{2}}{z}+\frac{ik}{z^{2}}-\frac{1}{z^{3}}) \, \CompVec{u}{x}$ (ch. 9.2. in \cite{jackson1999classical}). We find that the results of the simulations agree with the analytical expression to better than $\SI{2}{\%}$ (see Suppl. Figure S1). For $z \ll 1/k$, the $-1/z^{3}$ term dominates, and the scattered field is anti-parallel to the reference field. For $1/k \ll z$, the $1/z$ term dominates, and the scattered field is parallel to the reference field, in agreement with Figure \ref{fig3}b, \ref{fig3}c, and \ref{fig3}d. For $\lambda = \SI{532}{nm}$ and $n_{m}=1.33$, the transition occurs at  $z=1/k\approx \SI{64}{nm}$ (see Suppl. Figure S2).

The simulated Michelson contrast $C=(I_{max}-I_{min})/(I_{max}+I_{min})$ as a function of the distance to the screen (i.e. the photocathode) is shown in Figure \ref{fig3}e for different particle materials. The refractive indices of the protein and the virus were taken from~\cite{piliarik2014direct} and~\cite{ewers2007label}, respectively. For gold (Au) and copper (Cu), the complex refractive indices were interpolated at $\SI{532}{nm}$ from data in~\cite{johnson1972optical} and \cite{palik1998handbook}, respectively. In the near-field the contrast is found to drop as $C\propto 1/z^3$, which will enable excellent suppression of signal from scatterers that are not bound to the interface. 
Figure \ref{fig3}f shows the resolution of ONEM as a function of $z$, evaluated by the Full Width at Half Maximum (FWHM) of the central feature in the interference pattern along the $x$-direction. The calculation was again performed assuming a $R=\SI{2.5}{nm}$ protein. Within the plotted parameter range, the simulated FWHM increases linearly with the distance, with a fitted slope of $\partial FWHM / \partial z = 0.82$.

The expected contrast as a function of particle diameter is plotted in Figure 4 for three different materials. All of them are assumed to be immersed in water, and at a distance of $\SI{25}{nm}$ from the photocathode. In the absence of resonances, the polarizability is proportional to the volume of the particle, resulting in an $R^3$ dependence of the Michelson contrast. For metallic nanoparticles the contrast is seen to saturate for particle radii $>\SI{20}{nm}$.

\begin{figure}
  \includegraphics[width=7cm]{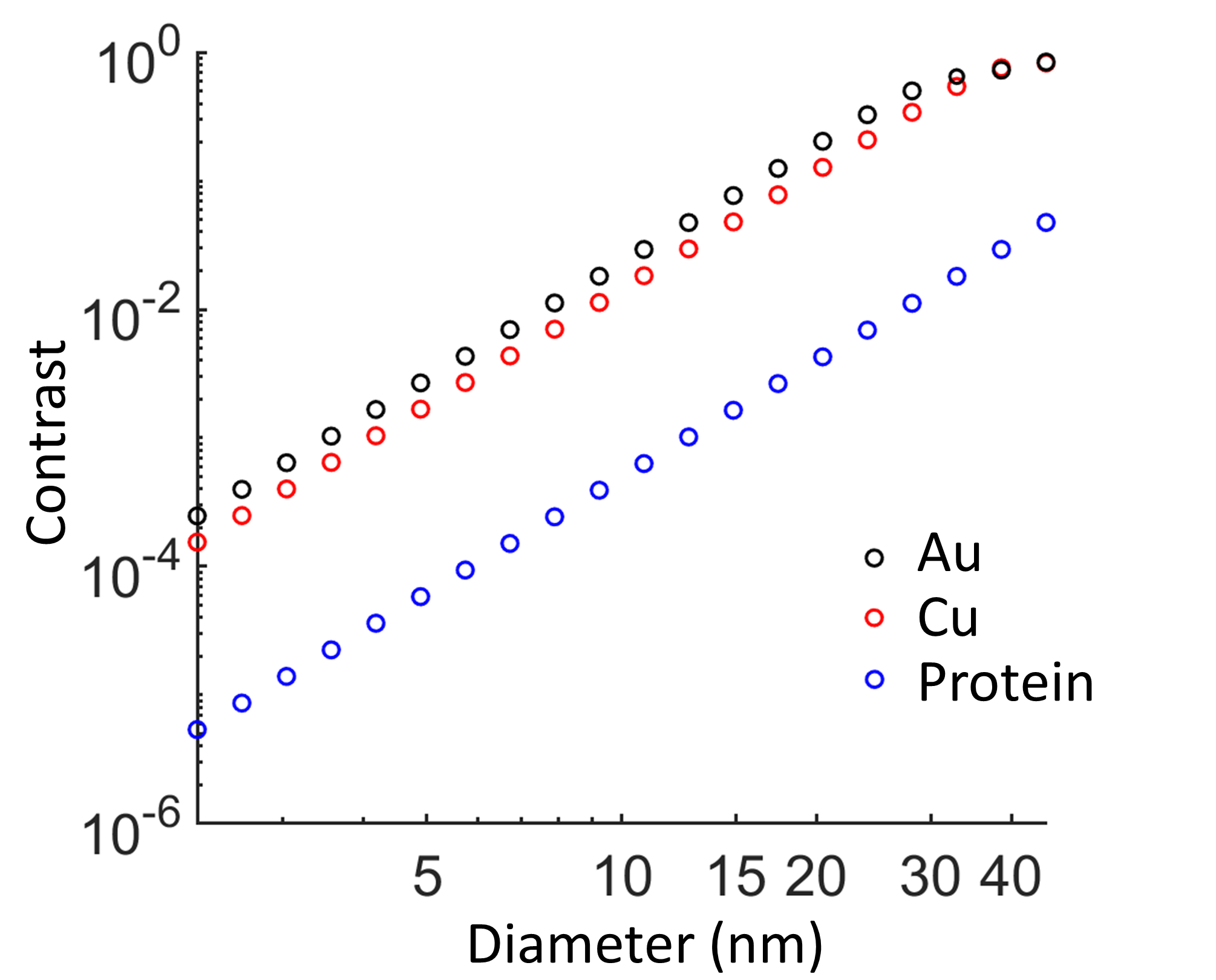}
  \caption{Michelson contrast of the simulated interference pattern $I(x, y)$ for a nanoparticle, as a function of the nanoparticle diameter, for $z=\SI{25}{nm}$.} 
  
  \label{fig4}
\end{figure}

\section{Potential application space}

By enabling dynamic studies of interfaces on the nanometer scale, ONEM has applications ranging from materials science to membrane biology.

One of them is the study of plasmonic fields. It is challenging to characterize light-matter interactions and devices based on optical near-fields, on the nanometer scale. Laser triggered electron microscopy has proven to be a versatile tool for mapping optical near-fields on nanometer spatial and femtosecond temporal scales~\cite{barwick2009photon, piazza2015simultaneous, kfir2020controlling}. ONEM could enable this in liquid environments, and at much lower optical excitation powers, allowing for a precise characterization, and consequent further development, of plasmonic (bio)sensors~\cite{gordon2019biosensing, taylor2013situ, oh2018performance} .  

ONEM could also be applied in electrochemistry, studying for example the nucleation and growth of nanoscale copper clusters in a liquid environment. Such experiments were the first application of \textit{in situ} liquid cell TEM~\cite{williamson2003dynamic}, starting a new area of research that has grown rapidly over the last decade~\cite{de2011electron}, both in material science and in biology. However, a serious and unresolved issue with such experiments is that the high-energy electron beam passes through the electrochemical cell above the working electrode, creating a large number of free, solvated electrons. These can strongly affect the electrochemistry~\cite{woehl2013experimental, Woehl2020}, and may even lead to the formation of hydrogen bubbles~\cite{Grogan2014}. ONEM could be used to observe the nucleation and growth during electrodeposition with nanometer resolution, and without electron-dose induced artefacts. Successful implementation of this prototype liquid cell experiment will open the door to investigating many electrochemical problems that are of significant industrial interest and that are currently out of reach for high-resolution, real-time studies (e.g. corrosion, mass transport in batteries, swelling, liquid crystal switching...). 

Finally, ONEM could also be used to image proteins interacting on, in, or with biomembranes under conditions mimicking a native membrane environment. It has been shown that lipid bilayers can be formed on surfaces with the use of cushions or tethers to allow for the embedding of proteins in the membrane, in particular transmembrane proteins, while maintaining the mobility and functionality of the inserted proteins ~\cite{machavn2010recent, roder2011reconstitution, poudel2011single}. If this is done on the vacuum-liquid interface, ONEM could be used for continuous high resolution imaging. First examples could include the visualization of supramolecular protein assemblies formed on lipid bilayers. In general, oligomerization of membrane proteins into multicomponent units is often critical for their function, or dysfunction. For instance, a change in the aggregation behavior of a pro-apoptotic protein Bax, subsequent formation of Bax complexes with a broad distribution of oligomerization numbers, and finally the opening of functional pores in the mitochondrial membrane represent key steps in programmed cell death~\cite{subburaj2015bax}.
Similarly, supra-molecular in-membrane assemblies consisting of $7-8$ monomeric units are formed by a functionally unrelated protein named fibroblast growth factor 2 (FGF2), which regulates tumor growth and metastasis. ONEM has the potential to provide information on the dimensions~\cite{steringer2017key, subburaj2015bax, antonsson2001bax} and also on the dynamics of these supra-molecular complexes, whose formation is crucial for cellular function. Additionally, since the contrast in ONEM decreases as $1/z^3$, unwanted protein signal from the solution is efficiently suppressed. It can therefore be used for imaging membrane related processes even in the cases where the equilibrium between bound and unbound protein is shifted in favour of the unbound protein.  

\section{Discussion and Outlook }

Optical Near-field Electron Microscopy (ONEM) represents a fundamentally new microscopy technique that exploits light and electron optics in an unprecedented way, and according to their respective strengths: light for non-invasive probing, electrons for high-spatial-resolution read-out. Light and electrons are coupled using an ultra-thin photocathode, and superresolution is enabled by the near-field components of the scattered light. ONEM is therefore ideally suited for the damage- and label-free study of physical, chemical, or biological processes happening at interfaces. A resolution below $\SI{5}{nm}$, and dynamic imaging over extended periods, seem feasible. Various experimental difficulties will have to be overcome to realize such specifications. A low energy electron microscope has to be modified to include a custom sample chamber, speckle-free optical excitation, as well as \textit{in situ} photocathode coating capabilities. Coating procedures will have to be optimized for sensitivity and homogeneity, and potential issues such as photocathode ageing, poisoning or charging will have to be addressed.  

ONEM can be operated, and extended, in many ways: Measurements can be performed in liquid, gas, or vacuum. They can be performed label-free, characterizing the refractive index distribution of a sample, or use labels (e.g. metal nanoparticles, fluorophores) bound to an object of interest. Correlative \textit{in situ} light microscopy can provide additional information, e.g. on the 3-dimensional environment of the interface, optionally with fluorescence enabled specificity. Polarization-dependent scattering cross-sections (e.g. chiral molecules or chiral photonic structures) could enable shot-noise limited measurements in challenging environments. Pump-probe measurements could make studies on ultra-short timescales possible. 

ONEM offers unique measurement opportunities on the dynamics of various processes, by allowing for the use of liquid cells: Optical (plasmonic) properties of nano-structured materials can be probed in a liquid environment, which can yield information that is essential for bio-optical sensor design. Electrochemical experiments will yield novel insights into physical and chemical processes at interfaces,  some of which play an important role in the energy transition. Finally, the exploration of tethered lipid bilayers will allow for the investigation of biological systems, offering information on protein clustering and dynamics within biological membranes.

Given recent developments in technology and methodology, we consider ONEM a feasible new form of microscopy. Clearly, several practical challenges are still to be overcome, but once these are addressed, ONEM will offer unique opportunities for damage-free imaging of dynamic processes at interfaces.

\medskip
\textbf{Acknowledgements} \par 

We thank Peter S. Neu for help with the figures.
This project has received funding from the European Union’s Horizon 2020 research and innovation programme under grant agreement No 101017902. TJ acknowledges support from the ERC MicroMOUPE Grant 758752. RM acknowledges funding from the European Union’s Framework Programme for Research and Innovation Horizon 2020 (2014‐2020) under the Marie Curie Skłodowska Grant Agreement Nr. 847548.
MA and MH acknowledge GAČR grant 19-26854X.

\medskip

\bibliographystyle{ieeetr}
\bibliography{ms}

\end{document}


\newcommand{\Vector}[1]{\vec{\textbf{\textit{#1}}}}

\newcommand{\CompVec}[2]{\vec{\textbf{\textit{#1}}}_{\textbf{\textit{#2}}}}

\widetext
\begin{center}
\textbf{\large Supplemental Materials for \\ \smallskip Optical Near-Field Electron Microscopy}

\end{center}

\setcounter{equation}{0}
\setcounter{figure}{0}
\setcounter{table}{0}
\setcounter{page}{1}
\makeatletter
\renewcommand{\theequation}{\textbf{S\arabic{equation}}}
\renewcommand{\thefigure}{S\arabic{figure}}
\renewcommand{\bibnumfmt}[1]{[#1]}
\renewcommand{\citenumfont}[1]{#1}

\section{On-axis scattered field: Comparison between analytical expression and numerical simulations}

Here we demonstrate the excellent agreement obtained between simulations using the MNPBEM toolbox \cite{hohenester2012mnpbem} and the analytic expression describing the field scattered by a dipole. Taking the expression of the x-polarized reference field from the main text, as well as the expression of the dipole moment $\Vector{p}(t)$, the polarizability $\alpha$, and the on-axis scattered field $\CompVec{E}{scat}(0, 0, z, t)$,  leads to:

\begin{equation}
    \CompVec{E}{scat}(0, 0, z, t)=\frac{3V}{4\pi} \frac{\epsilon_{1}-\epsilon_{m}}{\epsilon_{1}+2\epsilon_{m}} E_{ref} \textrm{e}^{i(kz-\omega t)}
    (\frac{k^{2}}{z}+\frac{ik}{z^{2}}-\frac{1}{z^{3}}) \, \CompVec{u}{x} \label{eq1}
\end{equation}

The amplitude of the on-axis scattered field, normalized to the amplitude of the reference field is therefore: 

\begin{equation}
    \frac{||\CompVec{E}{scat}(0, 0, z, t)||}{E_{ref}}=\frac{3V}{4\pi} \abs{ \frac{\epsilon_{1}-\epsilon_{m}}{\epsilon_{1}+2\epsilon_{m}}   
    (\frac{k^{2}}{z}+\frac{ik}{z^{2}}-\frac{1}{z^{3}})}\label{eq2}
\end{equation}

\begin{figure}[h]
  \includegraphics[width=14cm]{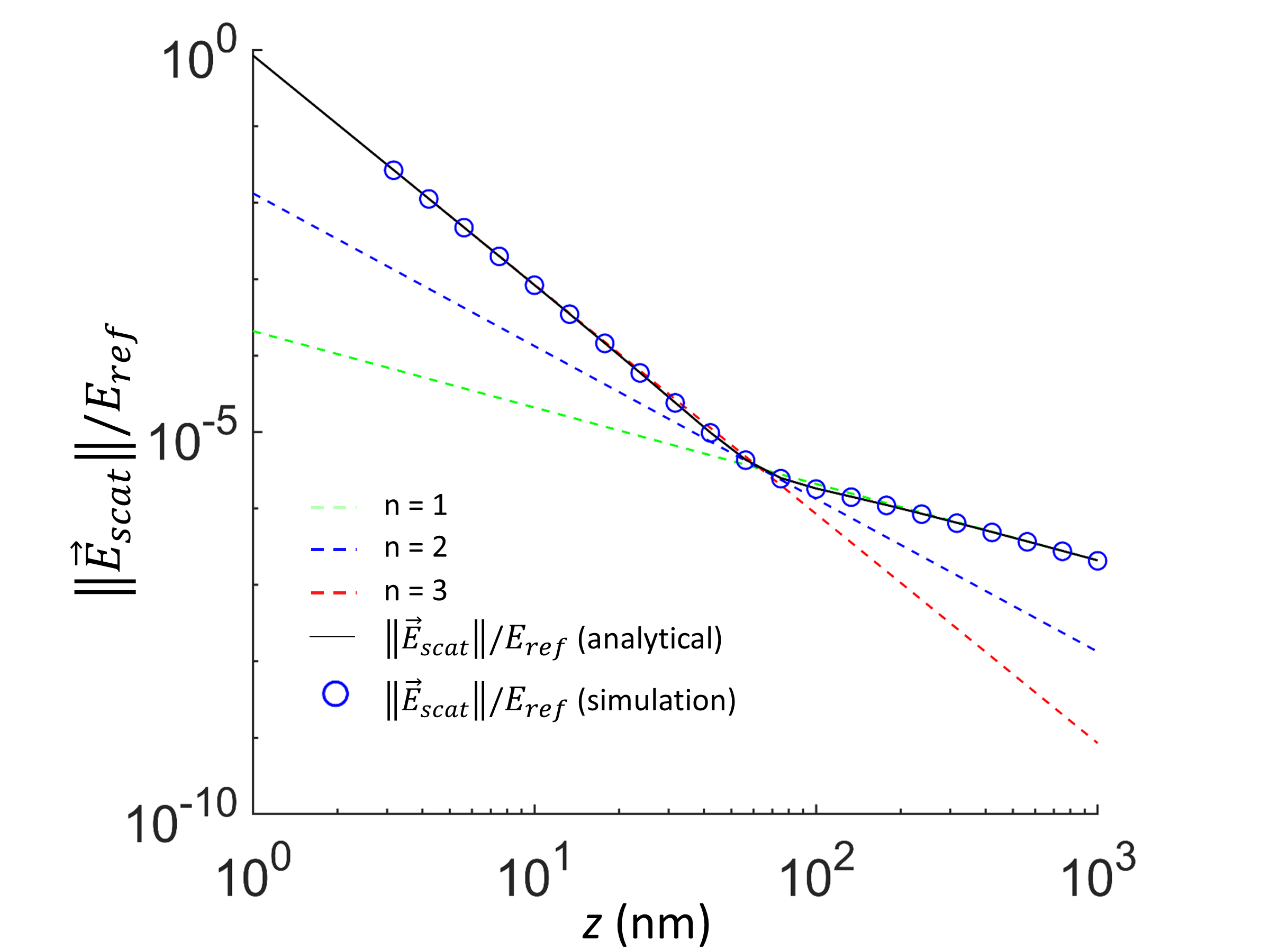}
  \caption{Amplitude of the on-axis scattered field for a spherical protein, normalized to the reference field: comparison between data simulated with the MNPBEM toolbox and the analytical expression \textbf{\ref{eq2}}. The dashed lines give the individual contributions of the terms $\propto 1/z^{n}$. The following parameters were used: $\epsilon_{1}=n_{1}^{2}$, $\epsilon_{m}=n_{m}^{2}$,  $n_{1}=\SI{1.44}{}$, and $n_{m}=\SI{1.33}{}$, $k=\frac{2 \, \pi \, n _{m}} {\lambda}$, $\lambda = \SI{532}{nm}$, $V=\frac{4}{3}\pi R^{3}$, $R=\SI{2.5}{nm}$. 
  }
  \label{figS1}
\end{figure}

Figure \ref{figS1} shows the excellent agreement between simulations and the analytical result from Eq. \textbf{\ref{eq2}}, in the case of a spherical protein. Note that the permittivities are real in this case. The terms $\propto 1/z^{n}$ are plotted to distinguish the near-field domain ($E_{scat}\propto z^{-3}$) and the far-field domain ($E_{scat}\propto z^{-1}$). The boundary between the near-field and the far-field domains occurs at $z\approx\SI{64}{nm}$.

\section{Interference pattern close to $z=1/k$}

\begin{figure}[h]
  \includegraphics[width=\linewidth]{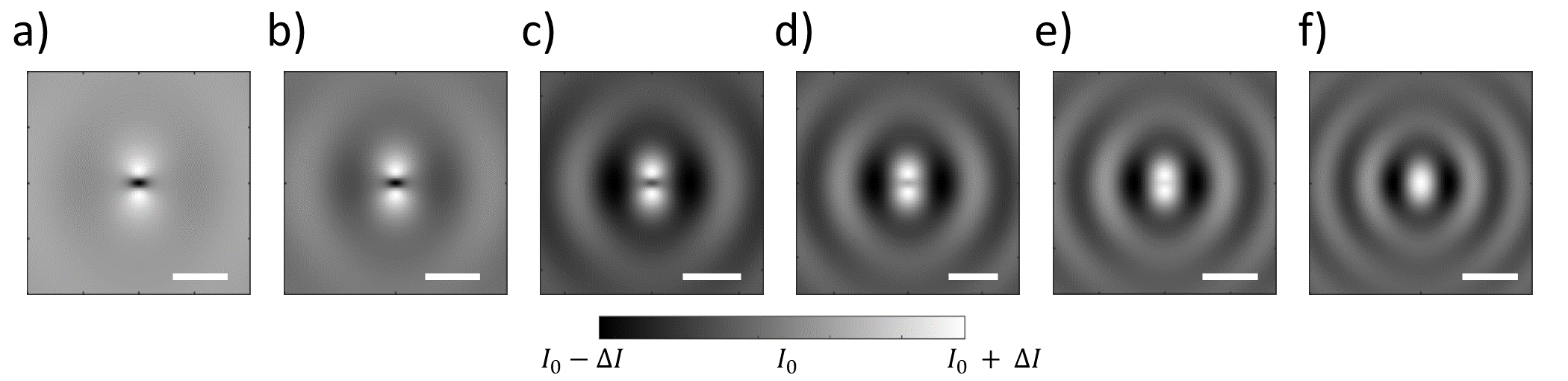}
  \caption{Simulated interference pattern $I(x, y)$ for a spherical protein of radius $R=\SI{2.5}{nm}$, and index of refraction $n_p=1.44$ at a) $z=\SI{40}{nm}$ (scale bar: \SI{200}{nm}, $\Delta I / I_0 = 1.3 \times 10^{-5} $), b) $z=\SI{50}{nm}$ (scale bar: \SI{250}{nm}, $\Delta I / I_0 = 5.6 \times 10^{-6} $), c) $z=\SI{64}{nm}$ (scale bar: \SI{300}{nm},  $\Delta I / I_0 = 2.7 \times 10^{-6} $), d) $z=\SI{75}{nm}$ (scale bar: \SI{300}{nm},  $\Delta I / I_0 = 2.3 \times 10^{-6} $), e) $z=\SI{85}{nm}$ (scale bar: \SI{400}{nm}, $\Delta I / I_0 = 2.1 \times 10^{-6} $), f) $z=\SI{100}{nm}$ (scale bar: \SI{500}{nm}, $\Delta I / I_0 = 2.0 \times 10^{-6} $).
  }
  \label{figS2}
\end{figure}

Figure \ref{figS2} shows the transition in the interference pattern $I(x, y)$ from $z < 1/k$ (a, b) to $z=1/k$ (c), and $z > 1/k$ (d, e, f). The simulations were done using the MNPBEM toolbox \cite{hohenester2012mnpbem}.

\bibliographystyle{ieeetr}

\bibliography{supplement}